\documentclass[conference,letterpaper,final,twocolumn]{IEEEtran}
\usepackage[utf8]{inputenc}
\usepackage{amsmath}
\usepackage{amssymb}
\usepackage{amsthm}
\newtheorem{theorem}{Theorem}

\newtheorem{lemma}{Lemma}

\newtheorem{remark}{Remark}
\title{An entropy inequality for $q$-ary random variables 
and its application to channel polarization}
\author{\IEEEauthorblockN{Eren \c Sa\c so\u glu}
\IEEEauthorblockA{EPFL, Lausanne, Switzerland\\eren.sasoglu@epfl.ch}}

\def\cX{\mathcal{X}}
\def\cY{\mathcal{Y}}

\def\uni{unif}
\begin{document}
\maketitle
\begin{abstract}
It is shown that given two copies of a $q$-ary input channel $W$,
where $q$ is prime, it is possible to create two channels
$W^-$ and $W^+$ whose symmetric capacities satisfy
$I(W^-)\le I(W)\le I(W^+)$, where the inequalities are strict except
in trivial cases. This leads to a simple proof of channel
polarization in the $q$-ary case.
\end{abstract}

\section{Introduction and Main Result}
Ar\i kan's \emph{polar codes}
\cite{Arikan2009} are a class of `symmetric capacity'-achieving codes
for binary-input channels. Their block
error probability behaves roughly like $O(2^{-\sqrt{N}})$
\cite{ArikanTelatar2009}, where $N$
is the blocklength, and they achieve this performance at an
encoding/decoding complexity of order $N\log N$. 

Polar codes for non-binary input channels were considered
in \cite{ArikanSasogluTelatar2009}. As in the binary case, their
construction is based on recursively creating new channels from
several copies of the original: Let $W$ be a discrete memoryless channel
with input alphabet $\cX=\{0,\dotsc,q-1\}$. Throughout this note,
$q$  will be assumed to be
a prime number. The output alphabet $\cY$ may be arbitrary. 
We will let $I(W)\in[0,1]$ denote
the mutual information developed across $W$ with uniformly distributed
inputs\footnote{All logarithms in this note will be to the base $q$.}, i.e.,
$$
I(W)=\sum_{x\in\cX,y\in\cY}\frac1qW(y\mid x)\log\frac{W(y\mid
x)}{\sum_{x'}\tfrac1qW(y\mid x')}.
$$
Let $X_1$, $X_2$ be independent, uniformly distributed inputs to two independent
copies of $W$, and let $Y_1$, $Y_2$ be the corresponding outputs.
Consider the one-to-one mapping $X_1,X_2\to
U_1,U_2$ 
\begin{align}
\label{eqn:transform}
\begin{split}
U_1&=X_1+X_2\\
U_2&=X_2,
\end{split}
\end{align}
where `$+$' denotes modulo-$q$ addition. Observe that $U_1$ and $U_2$ are
independent and uniformly distributed over $\cX$. Define the channels
\begin{align*}
W^-&\colon U_1\to Y_1Y_2,\\
W^+&\colon U_2\to Y_1Y_2U_1,
\end{align*}
described through the conditional output probability
distributions
\begin{align*}
W^-(y_1,y_2\mid u_1)&=\frac1q\sum_{u_2\in\cX} W(y_1\mid
u_1-u_2)W(y_2\mid u_2),\\
W^+(y_1,y_2,u_1\mid u_2)&=\frac1qW(y_1\mid u_1-u_2)W(y_2\mid u_2).\\
\end{align*}
It follows from the chain rule of mutual information that $I(W^-)+I(W^+)=2I(W)$. It is also easy to see that $W^+$ is
better than $W$, whereas $W^-$ is worse, in the sense that
\begin{align}
\label{eqn:worse-better}
I(W^-)\le I(W)\le I(W^+).
\end{align}
Since $W^-$ and $W^+$ are also $q$-ary input channels, the above
procedure can be applied to each of them, creating the channels
$W^{--}:=(W^-)^-$, $W^{-+}:=(W^-)^+$, $W^{+-}:=(W^+)^-$, and
$W^{++}:=(W^+)^+$. Repeating this procedure $n$ times, one obtains
$2^n$ channels, $W^\mathbf{s}$, $\mathbf{s}\in\{-,+\}^n$, with
$\sum_\mathbf{s}I(W^\mathbf{s})=2^nI(W)$. The main observation
that leads the author of \cite{Arikan2009} to construct polar codes is that these channels
are \emph{polarized} in the following sense:
\begin{theorem}[\cite{Arikan2009},\cite{ArikanSasogluTelatar2009}]
\label{thm:q-polarization}
$$
\lim_{n\to\infty}\frac1{2^n}\#\big\{\mathbf{s}\in\{-,+\}^n\colon
I(W^\mathbf{s})\in(1-\delta,1]\big\}=I(W),
$$
$$
\lim_{n\to\infty}\frac1{2^n}\#\big\{\mathbf{s}\in\{-,+\}^n\colon
I(W^\mathbf{s})\in[0,\delta)\big\}=1-I(W),
$$
for all $\delta>0$.
\end{theorem}
The proofs given in \cite{Arikan2009} and
\cite{ArikanSasogluTelatar2009} for Theorem~\ref{thm:q-polarization}
are based on the following arguments: The symmetric mutual informations of the
channels $W^{\mathbf{s}}$ created by the above procedure have a
martingale property, 
from which it follows that they must converge for almost all
paths in the construction.
This shows that both limits in
Theorem~\ref{thm:q-polarization} exist. To prove the claim on these
limits' values, it would be sufficient to show that
\eqref{eqn:worse-better} holds with strict inequalities 
for all $W^{\mathbf{s}}$, unless
$I(W^{\mathbf{s}})\in\{0,1\}$. Observe, however, that since the output
alphabets of channels $W^{\mathbf{s}}$ grow as the construction size increases,
this approach would require the aforementioned inequality to hold
uniformly for all $q$-ary input channels. This difficulty is
circumvented in \cite{Arikan2009} and \cite{ArikanSasogluTelatar2009} 
by appropriately defining an auxiliary channel parameter $Z(W)$ and proving the
convergence of $Z(W^{\mathbf{s}})$ to $\{0,1\}$ by the above
arguments, which then implies the convergence of $I(W^\mathbf{s})$ to
$\{0,1\}$. 

The purpose of this note is to provide a proof of
Theorem~\ref{thm:q-polarization} that avoids this indirect approach.
In order to do so, we will need the following theorem.
\begin{theorem}
\label{thm:main}
If $I(W)\in(\delta,1-\delta)$ for some $\delta>0$, then there
exists an $\epsilon(\delta)>0$ such that
$$
I(W^-)+\epsilon(\delta)\le I(W)\le I(W^+)-\epsilon(\delta).
$$
The dependence of $\epsilon(\delta)$ on the channel $W$ is only
through $\delta$, and not through particular channel specifications
(e.g., output alphabet size). 
\end{theorem}
Theorem~\ref{thm:main} will be proved as a corollary to the following
lemma, which is the main result reported here.
\begin{lemma}
\label{lem:entropy}
Let $X_1,X_2\in\cX$, $Y_1,Y_2\in\cY$ be random variables with joint
probability density
\begin{align}
\label{eqn:density}
\begin{split}
P_{X_1Y_1X_2Y_2}(x_1,y_1,&x_2,y_2)\\
&\hfill=P_{X_1Y_1}(x_1,y_1)P_{X_2Y_2}(x_2,y_2).
\end{split}
\end{align}
If 
$$
H(X_1\mid Y_1),H(X_2\mid Y_2)\in(\delta,1-\delta)
$$ 
for some $\delta>0$, then there exists an
$\epsilon(\delta)>0$ such that
$$
H(X_1+X_2\mid Y_1,Y_2)-\max\{H(X_1\mid Y_1),H(X_2\mid Y_2)\}\ge\epsilon(\delta).
$$
\end{lemma}
We will prove Lemma~\ref{lem:entropy} in Section~\ref{sec:proof}.

\begin{proof}[Proof of Theorem~\ref{thm:main}]
It suffices to show that $I(W)-I(W^-)\ge\epsilon(\delta)$, as the
equality $I(W^-)+I(W^+)=2I(W)$ will then imply the second half of the claim.
Let $X_1,X_2\in\cX$ denote two independent and uniformly distributed
inputs to two copies of $W$, and let $Y_1,Y_2\in\cY$ be the
corresponding outputs. Since $W$ is memoryless, $X_1,X_2,Y_1,Y_2$ are
jointly distributed as in \eqref{eqn:density}.
Further,
$I(W)\in(\delta,1-\delta)$ implies
\begin{align}
\label{eqn:I(W)}
1-I(W)=H(X_1\mid Y_1)=H(X_2\mid Y_2)\in(\delta,1-\delta).
\end{align}
It then follows from Lemma~\ref{lem:entropy} that
\begin{align*}
I(W)-I(W^-)&=H(X_1+X_2\mid Y_1Y_2)-H(X_1\mid Y_1)\\
&\ge\epsilon(\delta),
\end{align*}
completing the proof.
\end{proof}

\section{Proof of Theorem~\ref{thm:q-polarization}}
\label{sec:polarization}
Let $B_1,B_2,\dotsc$ be $\{-,+\}$-valued i.i.d.\ random variables
with $\Pr[B_1=-]=\Pr[B_1=+]=\tfrac12$. Let $I_0,I_1,\dotsc$ be random
variables defined as
\begin{align*}
I_0&=I(W)\\
I_n&=I(W^{B_1,\dotsc,B_n})\quad n=1,2,\dotsc
\end{align*}
Note that $I_n$ takes values in $[0,1]$. Further, 
it follows from the relation $I(W^-)+I(W^+)=2I(W)$ that
$\mathbb{E}[I_{n+1}\mid I_n,\dotsc,I_0]=I_n$. Hence, the process $I_0,I_1,\dotsc$ is
a bounded martingale, and therefore converges almost surely to a $[0,1]$-valued random variable $I_\infty$. 
Note, on the other hand, that
$$
\Pr[I_n\in(\delta,1-\delta)]=\frac1{2^n}\#\big\{\mathbf{s}\in\{-,+\}^n\colon
I(W^\mathbf{s})\in(\delta,1-\delta)\big\}.
$$
To conclude the proof, it thus suffices to show that
$\Pr[I_\infty=1]=I(W)$ and $\Pr[I_\infty=0]=1-I(W)$. To that end, note that the
almost sure convergence of $I_n$ implies
$\mathbb{E}[|I_{n+1}-I_n|]=\mathbb{E}[I(W^{B_1\dotsc B_n+})-I(W^{B_1\dotsc
B_n})]\to0$. It follows from Theorem~\ref{thm:main}
that the latter convergence implies
$I_\infty\in\{0,1\}$ with probability $1$. Due to the
martingale property of $I_n$ we have
$\mathbb{E}[I_\infty]=\mathbb{E}[I_0]=I(W)$, from which it follows
that $\Pr[I_\infty=1]=1-\Pr[I_\infty=0]=I(W)$, completing the proof.

\section{Proof of Lemma~\ref{lem:entropy}}
\label{sec:proof}
In what follows, $H(p)$ and $H(X)$ will both denote the entropy of a
random variable $X\in\cX$ with probability distribution
$p$. We will let $p_i$, $i\in\cX$ denote the probability
distribution with
$$
p_i(m)=p(m-i).
$$
The cyclic convolution of vectors $p$ and $r$ will be
denoted by $(p\ast r)$. That is,
$$
(p*r)=\sum_{i\in\cX}p(i)r_i=\sum_{i\in\cX}r(i)p_i.
$$
We will also let $\uni(\cX)$ denote the uniform
distribution over $\cX$. 
We will use the following lemmas in the proof:

\begin{lemma}
\label{lem:variation-divergence}
Let $p$ be a distribution over $\cX$. Then,
$$
\|p-\uni(\mathcal{X})\|_1\ge \frac{1}{q\log
e}[1-H(p)] .
$$
\end{lemma}
\begin{remark}
Lemma~\ref{lem:variation-divergence} partially complements
Pinsker's inequality by providing a lower bound to the
$\mathcal{L}_1$ distance between an arbitrary probability distribution
and the uniform
distribution by their Kullback--Leibler divergence.
\end{remark}
\begin{proof}
\begin{align*}
1-H(p)&=\sum_{i\in\cX} p(i)\log\frac{p(i)}{1/q}\\
&\le\log e \sum_i p(i)\left[ \frac{p(i)-1/q}{1/q}\right]\\
&\le q\log e \sum_i p(i)|p(i)-1/q|\\
&\le q\log e\|p-\uni(\mathcal{X})\|_1,
\end{align*}
where we used the relation $\ln t\le t-1$ in the first inequality.
\end{proof}
\begin{remark}
Lemma~\ref{lem:variation-divergence} holds for distributions over
arbitrary finite sets. That $|\cX|$ is a prime number
has no bearing on the above proof.
\end{remark}

\begin{lemma}
\label{lem:shift-distance}
Let $p$ be a distribution over $\cX$. Then,
$$
\|p_i-p_j\|_1\ge \frac{1-H(p)}{2q^2(q-1)\log e} .
$$
for all $i,j\in\cX$, $i\neq j$. That is, unless $p$ is the uniform
distribution, its cyclic shifts will be separated from each other in
the $\mathcal{L}_1$ distance.
\end{lemma}
\begin{proof}
Let $j=i+m$ for some $m\neq0$. 
We will show that there exists a $k\in\cX$
satisfying
$$
|p(k)-p(k+m)|\ge\frac{1-H(p)}{2q^2(q-1)\log e},
$$
which will yield the claim since
$\|p_i-p_j\|_1=\sum_{k\in\cX}|p(k)-p(k+m)|$.

Suppose that $H(p)<1$, as the claim is trivial otherwise. Let
$p^{(\ell)}$ denote the $\ell$th largest element of $p$, 
and let $S=\{\ell:p^{(\ell)}\ge\tfrac1q\}$. Note that $S$ is a proper subset
of $\cX$.
We have
\begin{align*}
\sum_{\ell=1}^{|S|}[p^{(\ell)}-p^{(\ell+1)}]&=p^{(1)}-p^{(|S|+1)}\\
&\ge p^{(1)}-1/q\\
&\ge \frac1{2(q-1)}\|p-\uni(\cX)\|_1\\
&\ge \frac{1-H(p)}{2q(q-1)\log e}.
\end{align*}
In the above, the second inequality is obtained by observing that
$p^{(1)}-1/q$ is smallest when $p^{(1)}=\dots=p^{(q-1)}$, and the third inequality follows from
Lemma~\ref{lem:variation-divergence}. Therefore, there exists at least
one $\ell\in S$ such that
$$
p^{(\ell)}-p^{(\ell+1)}\ge \frac{1-H(p)}{2q^2(q-1)\log e} .
$$
Given such an $\ell$, let $A=\{1,\dotsc,\ell\}$. Since $q$ is prime, $\cX$
can be written as 
$$
\cX=\{k,k+m,k+m+m,\dotsc,k\underbrace{+m+\dotsc+m}_{q-1\text{ times}}\}
$$
for any $k\in\cX$ and $m\in\cX\backslash\{0\}$. Therefore,
since $A$ is a proper subset of $\cX$, there exists a
$k\in A$ such that
$k+m\in A^c$, implying
$$
p(k)-p(k+m)\ge \frac{1-H(p)}{2q^2(q-1) \log e},
$$
which yields the claim.
\end{proof}

\begin{lemma}
\label{lem:H-convolution}
Let $p$ and $r$ be two probability distributions over $\cX$, with
$H(p)\ge\eta$ and $H(r)\le1-\eta$ for some $\eta>0$. Then, there
exists an $\epsilon_1(\eta)>0$ such that
$$
H(p\ast r)\ge H(r)+\epsilon_1(\eta).
$$
\end{lemma}
\begin{proof}
Let $e_i$ denote the distribution with a unit mass on $i\in\cX$. Since
$H(p)\ge\eta>H(e_i)=0$, it
follows from the continuity of entropy that
\begin{align}
\label{eqn:interior}
\min_i \|p-e_i\|_1\ge \mu(\eta)
\end{align}
for some $\mu(\eta)>0$.
On the other hand, since $H(r)\le1-\eta$, we have by
Lemma~\ref{lem:shift-distance} that
\begin{align}
\label{eqn:distance}
\|r_i-r_j\|_1\ge \frac{\eta}{2q^2(q-1)\log e}>0
\end{align}
for all pairs $i\neq j$. Relations \eqref{eqn:interior},
\eqref{eqn:distance}, and the strict concavity of entropy implies
the existence of $\epsilon_1(\eta)>0$ such that
\begin{align*}
H(p\ast r)&=H\left(\sum_i p(i)r_i\right)\\&\ge\sum_i p(i)H(r_i) +
\epsilon_1(\eta)\\&=H(r)+\epsilon_1(\eta).
\end{align*}
\end{proof}

\begin{proof}[Proof of Lemma~\ref{lem:entropy}]
Let $P_1$ and $P_2$ be two random probability
distributions on $\cX$,
with
\begin{align*}
P_1=P_{X_1\mid Y_1}(\cdot\mid y_1)\text{ whenever } Y_1=y_1,\\
P_2=P_{X_2\mid Y_2}(\cdot\mid y_2)\text{ whenever } Y_2=y_2.
\end{align*}
It is then easy to see that 
\begin{align*}
H(X_1\mid Y_1)&=\mathbb{E}[H(P_1)],\\
H(X_2\mid Y_2)&=\mathbb{E}[H(P_2)],\\
H(X_1+X_2\mid Y_1,Y_2)&=\mathbb{E}[H(P_1\ast P_2)].
\end{align*}
Suppose, without loss of generality, that $H(X_1\mid Y_1)\ge H(X_2\mid
Y_2)$. 
It suffices to show that if
$\mathbb{E}[H(P_1)],\mathbb{E}[H(P_2)]\in(\delta,1-\delta)$ for some $\delta>0$, then
there exists an $\epsilon(\delta)>0$ such that $\mathbb{E}[H(P_1\ast
P_2)]\ge\mathbb{E}[H(P_1)]+\epsilon(\delta)$. To that end, define  the
event
$$
A=\{H(P_1)>\delta/2,\hspace{.2em}H(P_2)<1-\delta/2\}.
$$
Observe that
\begin{align*}
\delta&<\mathbb{E}[H(P_1)]\\ &\le
\big(1-\Pr[H(P_1)>\delta/2]\big)\cdot\delta/2+\Pr[H(P_1)>\delta/2],
\end{align*}
implying $\Pr[H(P_1)>\delta/2]>\tfrac{\delta}{2-\delta}$. It similarly
follows that $\Pr[H(P_2)<1-\delta/2]>\tfrac{\delta}{2-\delta}$. Note
further that $H(P_1)$ and $H(P_2)$ are independent since $Y_1$ and
$Y_2$ are. Thus,
$A$ has probability at least
$\frac{\delta^2}{(2-\delta)^2}=:\epsilon_2(\delta)$. On the other
hand, Lemma~\ref{lem:H-convolution} implies that conditioned on 
$A$ we have
\begin{align}
\label{eqn:A}
H(P_1\ast P_2)\ge H(P_1)+\epsilon_1(\delta/2)
\end{align}
for some $\epsilon_1(\delta/2)>0$. Thus,
\begin{align*}
&\mathbb{E}[H(P_1\ast P_2)]\\&=\Pr[A]\cdot\mathbb{E}[H(P_1\ast P_2)\mid
A]+\Pr[A^c]\cdot\mathbb{E}[H(P_1\ast P_2)\mid A^c]\\
&\ge\Pr[A]\cdot \mathbb{E}[\big(H(P_1)+\epsilon_1(\delta/2)\big)\mid
A]\\
&\hspace{10em}+\Pr[A^c]\cdot\mathbb{E}[H(P_1)\mid A^c]\\
&\ge \mathbb{E}[H(P_1)]+\epsilon_1(\delta/2)\epsilon_2(\delta),
\end{align*}
where in the first inequality we used \eqref{eqn:A} and the relation
$H(p\ast r)\ge H(p)$. Setting $\epsilon(\delta):=\epsilon_1(\delta/2)\epsilon_2(\delta)$
yields the result.
\end{proof}

\section{Discussion}
The proof of Theorem~\ref{thm:main} does not extend trivially to the
case of composite input alphabet sizes. In particular, that the cyclic
group $\big(\{0,\dotsc,q-1\},+\big)$ is generated by each of its
non-zero elements is crucial to the proof of
Lemma~\ref{lem:shift-distance}. On the
other hand, a weaker statement holds when the input alphabet size is
composite: Consider replacing the mapping \eqref{eqn:transform} with
\begin{align}
\label{eqn:transform2}
\begin{split}
U_1&=X_1+X_2,\\
U_2&=\pi(X_2),
\end{split}
\end{align}
where $\pi$ is a permutation over $\cX$, and define the channels
$W^-\colon U_1\to Y_1Y_2$ and $W^+\colon U_2\to Y_1Y_2U_1$ accordingly. Then, it
can be shown that there exists a permutation $\pi$ for which
Theorem~\ref{thm:main} holds, irrespective of the input alphabet size.
The proof of this statement is similar to that of
Theorem~\ref{thm:main}, and therefore is omitted. It then follows
that channels with composite 
input alphabet sizes can be polarized in the sense 
\vfill
\pagebreak
\noindent
of Theorem~\ref{thm:q-polarization} if the mapping in \eqref{eqn:transform2}
is chosen appropriately at each step of construction. Whether
such channels can be polarized by recursive application of a
\emph{fixed} mapping is an open question.
\section*{Acknowledgment}
I would like to thank Emre Telatar for helpful discussions.

\end{document}